\documentclass[fleqn,usenatbib]{mnras}
\usepackage{newtxtext,newtxmath}

\usepackage[T1]{fontenc}
\usepackage{ae,aecompl}

\usepackage{graphicx}    
\usepackage{amsmath}    

\defcitealias{HM20}{HM20}
\defcitealias{Hirashita/Aoyama:2019}{HA19}
\defcitealias{Huang/etal:2020}{H20}
\defcitealias{MRN}{MRN}

\title[Cosmic grain size evolution]{Cosmic evolution of grain size distribution in galaxies using the $\nu^2$GC semi-analytic model}

\author[R. Makiya et al.]{
Ryu Makiya$^{1,2}$\thanks{E-mail: rmakiya@asiaa.sinica.edu.tw}
and Hiroyuki Hirashita$^{1}$\thanks{E-mail: hirashita@asiaa.sinica.edu.tw}
\\
$^{1}$Institute of Astronomy and Astrophysics, Academia Sinica,
Astronomy-Mathematics Building, No.\ 1, Section 4,
Roosevelt Road, Taipei 10617, Taiwan\\
$^{2}$Kavli Institute for the Physics and Mathematics of the Universe (Kavli IPMU, WPI), Todai Institutes for Advanced Study,
the University of Tokyo,\\ Kashiwa 277-8583, Japan
}

\date{Accepted XXX. Received YYY; in original form ZZZ}

\pubyear{2022}

\begin{document}
\label{firstpage}
\pagerange{\pageref{firstpage}--\pageref{lastpage}}
\maketitle

\begin{abstract}
We investigate the cosmological evolution of interstellar dust with a semi-analytical galaxy formation model ($\nu^2$GC), focusing on the evolution of grain size distribution.
The model predicts the statistical properties of dust mass and grain size distribution in galaxies across cosmic history.
We confirm that the model reproduces the relation between dust-to-gas ratio and metallicity in the local Universe, and that the grain size distributions of the Milky Way (MW)-like sample become similar to the so-called MRN distribution that reproduces the observed MW extinction curve.
Our model, however, tends to overpredict the dust mass function at the massive end at redshift $z\lesssim 0.8$ while it reproduces the abundance of dusty galaxies at higher redshifts.
We also examine the correlation between grain size distribution and galaxy properties (metallicity, specific star formation rate, gas fraction, and stellar mass), and observe a clear trend of large-grain-dominated, small-grain-dominated, and MRN-like grain size distributions from unevolved to evolved stages. 
As a consequence, the extinction curve shapes are flat, steep, and intermediate (MW-like) from the unevolved to evolved phases. At a fixed metallicity, the grain size distribution tends to have larger fractions of small grains at lower redshift; accordingly, the extinction curve tends to be steeper at lower redshift. 
We also predict that supersolar-metallicity objects at high redshift have flat extinction curves with weak 2175 \AA\ bump strength.
\end{abstract}

\begin{keywords}
dust, extinction -- Galaxy: evolution -- galaxies: evolution -- galaxies: high-redshift -- galaxies: ISM.
\end{keywords}


\section{Introduction}
\label{sec:intro}
Dust plays an important role in galaxy evolution by, for example, depleting the gas-phase metals \citep[e.g.][]{Jenkins:2009}, forming molecules (especially H$_2$) on dust surfaces \citep[e.g.][]{Cazaux:2004}, and reprocessing ultraviolet (UV)--optical radiation into the infrared (IR) \citep[e.g.][]{Draine/Li:2007}.
In particular, the last process shapes the observed spectral energy distributions (SEDs) of galaxies  \citep[e.g.][]{Takagi:2003,Nishida:2022}.
Indeed, various models that aim at extracting physical properties of galaxies such as star formation rate (SFR), stellar mass, etc., take dust extinction (absorption plus scattering) and reemission into account \citep[e.g.][]{daCunha:2008,Boquien:2019,Fioc:2019,Abdurrouf:2021}.
Thus, the understanding of dust evolution provides an essential key to how galaxy properties evolve in the Universe.

There have been some efforts of constructing theoretical models for dust production and evolution in a consistent manner with galaxy formation and evolution. 
A viable approach is to implement dust evolution models in cosmological hydrodynamic simulations \citep[e.g.][]{McKinnon:2016,Aoyama/etal:2018,Hou:2019}, which are able to calculate the gravitational evolution of dark matter and baryons together with baryonic processes (e.g.\ radiative and hydrodynamic processes) in formed galaxies.
In these simulations, dust enrichment by stars [mainly supernovae (SNe) and asymptotic giant branch (AGB) stars] and dust loss by star formation (astration), SN shock destruction, and outflow are considered in a manner consistent with the star formation and stellar feedback predicted from hydrodynamic evolution of baryons. As a consequence, these simulations have provided useful tools to understand statistical properties of dust (dust mass, dust abundance, etc.) in galaxies.

Most simulations focus on the total dust mass or the abundances of various dust species \citep[e.g.][]{Choban/etal:2022}. 
However, many of the processes in which dust plays an important role are also regulated by the grain size distribution (the distribution function of dust grain size).
For example, the formation rate of H$_2$ through the dust-surface reaction is proportional to the total surface area, which depends on the grain size distribution even if the total dust abundance is fixed \citep{Yamasawa:2011,Chen:2018,Romano/etal:2022}. 
The extinction of stellar light also depends on the grain size distribution, which governs the wavelength dependence of extinction \citep[the extinction curve; e.g.][hereafter MRN]{MRN}. 
Therefore, the evolution of grain size distribution is of fundamental importance in understanding the influence of dust on galaxy properties.

Some hydrodynamic simulations of galaxies incorporated the evolution of grain size distribution.
\citet{McKinnon:2018} developed a scheme for calculating the grain size distribution, treating the dust as a separate component from the gas. 
They presented some test calculation results for isolated galaxies, showing that newly included processes such as shattering (grain disruption) and coagulation (grain--grain sticking) indeed play a dominant role in changing the grain size distribution. 
\citet{Aoyama:2020} implemented the evolution of grain size distribution in a hydrodynamic simulation of isolated galaxy and found that the grain size distribution depends on the physical condition (especially the density) of the ambient gas, which affects the efficiencies of shattering and coagulation.
\cite{Romano/etal:2022} additionally included the effect of diffusion (exchange of the medium among adjacent hydro elements) and showed that diffusion not only homogenizes the grain size distribution in different interstellar medium (ISM) phases but also affects the overall grain size distribution by enhancing the production of small grains through shattering and accretion. 
All these three simulations only focused on a single galaxy for their runs. 
A cosmological simulation with a full implementation of grain size distribution was first performed by \citet{Li:2021}, who focused on the Milky Way (MW)-like galaxies and showed that their predicted extinction curves are consistent with the observed MW extinction curve, albeit a large variation. 
\citet[][hereafter H20]{Huang/etal:2020} took a different approach by post-processing a cosmological hydrodynamic simulation with their evolution model of grain size distribution, originally developed by \citet{Asano/etal:2013} and \citet[][hereafter HA19]{Hirashita/Aoyama:2019}, and focused on MW analogs.
They also broadly reproduced the MW extinction curve, but still found the importance of subgrid treatment for shattering and coagulation. 
Both of these two studies on the cosmological evolution of grain size distribution focused on the MW-like galaxies, but did not expand their predictions to general galaxy populations, mainly because of high computational cost required for the calculation of grain size distribution.

There are some solutions to reduce the computational cost for cosmological calculations of grain size distribution. 
One is to simplify the grain size distribution by representing the entire grain radius range with two sizes (`large' and `small' grains) as proposed by \cite{Hirashita:2015}. 
The implementation of this `two size approximation' into a cosmological hydrodynamic simulation was successfully performed by e.g.\ \citet{Aoyama/etal:2018}, \citet{Hou:2019}, \citet{Gjergo:2018}, and \citet{Granato:2021}.
The largest disadvantage of this approach is that there is only two degrees of freedom in the extinction curve (for each dust component if multiple species are included).
Another way of saving the computational cost is to use semi-analytic galaxy evolution models \citep[see, e.g.][for review]{Somerville/Dave:2015} instead of hydrodynamic simulations.
In semi-analytic models, merger trees of dark matter halos are generated by a $N$-body simulation or an analytic treatment, and galaxies are assigned to each dark matter halo. 
Galaxy evolution is treated along the merger trees by adopting simplified but physically motivated recipes of complicated baryonic processes, such as gas cooling, star formation and feedback, metal enrichment, and galaxy merger.

There have been some semi-analytic models in which dust evolution is implemented.
In most of them, the dust abundance is estimated by using empirical relations.
For example, the \textsc{galform} semi-analytic model \citep{Lacey/etal:2016} assumed the dust-to-gas ratio to be simply proportional to the metallicity. 
In the \textsc{shark} semi-analytic model, \cite{Lagos/etal:2019} employed the observed relation between gas-phase metallicity and dust-to-gas ratio of local galaxies obtained by \cite{RemyRuyer/etal:2014} to predict the dust mass in each galaxy.
Dust evolution calculations are implemented in the Santa Cruz semi-analytic model \citep{Popping/etal:2017}, \textsc{L-Galaxies} \citep{Vijayan/etal:2019}, and \textsc{sage} \citep{Triani:2020,Triani:2021}, all of which track the physical processes related to dust formation in stellar sources, and dust growth and destruction in the ISM.
The \textsc{gamete} semi-analytic model also treated these dust evolution processes, and focused on the assembly of the MW \citep{de-Bennassuti:2014,Ginolfi:2018}.
All the above semi-analytic models are capable of predicting statistical properties of dust abundance in galaxies, such as dust mass functions, the relation between dust-to-gas ratio and metallicity, etc., across cosmic history. 
However, there has been no semi-analytic models that successfully calculate the evolution of grain size distribution. 
Therefore, the dust-related quantities strongly affected by the grain size distribution are not predicted well, which has limited our understanding of, for example, the statistical properties of extinction curves in the entire galaxy population at various redshifts.

In this paper we study the cosmological evolution of dust mass and grain size distribution in galaxies by combining our dust model, \citetalias{Hirashita/Aoyama:2019} and \citet[][hereafter HM20]{HM20}, and the New Numerical Galaxy Catalogue ($\nu^2$GC) semi-analytic galaxy formation model \citep{Nagashima/etal:2005,Ishiyama/etal:2015,Makiya/etal:2016,Shirakata/etal:2019,Oogi/etal:2020}.
In this way, we are able to investigate the statistical properties of grain size distribution in the cosmic volume across cosmic history. 
As done in previous studies, we predict the extinction curve as an observable property related to the grain size distribution. 
This step provides a critical check against the actually observed extinction curves in nearby galaxies. 
The prediction on the dust abundance is also utilized to test if the model successfully reproduces the statistical properties of the total dust budget in the Universe.

The paper is structured as follows.
In Section \ref{sec:model} we briefly describe the $\nu^2$GC galaxy formation model and the dust evolution model.
In Section \ref{sec:results_dust}, we show basic results and comparison with some available observations. In Section \ref{sec:prediction}, we discuss the relation between grain size distribution and galaxy properties.
In Section \ref{sec:summary}, we provide the summary and conclusions of this paper.
We assume the flat-$\Lambda$CDM cosmology and adopt the cosmological parameters of {\it Planck} 2014 \citep{Planck:2014} throughout this work, namely, $\Omega_0$ = 0.31,
$\Omega_b$ = 0.048, $h$ = 0.68, $n_s$ = 0.96, and $\sigma_8$ = 0.83.

\section{Model}
\label{sec:model}
\subsection{Overview of $\nu^2$GC}
\label{subsec:nu2GC}
We use the $\nu^2$GC semi-analytic model.
The merger trees of dark matter (DM) halos are constructed from a cosmological $N$-body simulation \citep{Ishiyama/etal:2015}. 
The $\nu^2$GC $N$-body simulations, which consist of six variations with varying box sizes and mass resolutions, are performed by a massively parallel TreePM code \textsc{GreeM} \citep{Ishiyama/etal:2009,Ishiyama/etal:2012}.
In this paper we use the $\nu^2$GC-SS simulation, which has a box size of 70 $h^{-1}{\rm Mpc}$ and a minimum halo mass of $8.79\times10^9$ $h^{-1}M_{\odot}$.
The DM halos are detected by the friends-of-friends algorithm \citep{Davis/etal:1985} with the linking parameter of $b = 0.2$ and a minimum particle number of 40.

Based on the constructed merger trees, the semi-analytic galaxy evolution model developed by \citet{Nagashima/etal:2005} and later modified by \citet{Makiya/etal:2016} is applied for physical processes of baryonic components (gas and stars).
In the model, gas is first shock heated when the associated DM halo collapses and then it is cooled by radiative cooling to eventually form a cold gas disc.
Stars are formed from the cooled gas on a time-scale proportional to the dynamical time of the disc.
The metal enrichment of the ISM is calculated in a manner consistent with star formation history assuming a canonical initial mass function \citep{Kroupa:2001} as it is prescribed in \citet{Chabrier:2003}.
Stellar feedback in the form of energy input from SNe is taken into account by converting the cold gas into the hot gas phase.
Galaxy merging associated with the merger of DM halos is also considered, forming a bulge component.
The merger events also induce gas accretion onto the central supermassive black holes (SMBHs), maintaining the tight bulge--SMBH mass correlation.
The energy input from the central SMBH is also considered in the form of the suppression of star formation activity.
This semi-analytic model as successful as other models: It reproduces
the cosmic star formation history within the uncertainties in the observational data, as well as the stellar mass function and cold gas mass function in the local Universe.
It also broadly reproduces the stellar mass--metalicity relation of local star-forming galaxies, although there is a slight tendency of overpredicting metallicity for massive galaxies.
The detailed description of the $\nu^2$GC model is given by \citet{Ishiyama/etal:2015}, \citet{Makiya/etal:2016}, and \citet{Shirakata/etal:2019}.

The current version of $\nu^2$GC does not explicitly model the evolution of interstellar dust but only calculates the dust extinction of stellar light by assuming that the dust mass is proportional to the metallicity and cold gas mass.
In this paper, we calculate the physical evolution of dust mass and grain size distribution by post-processing the mock galaxy catalog of $\nu^2$GC using our dust evolution model (Section \ref{subsec:dustmodel}).
The key inputs for this dust evolution model are the cold gas mass fraction and the gas-phase metallicity. 
Thus, it is important that the $\nu^2$GC model reproduces the local cold gas mass function and the stellar mass--metallicity relation \citep{Makiya/etal:2016}.
This semi-analytic model is used to predict the grain size distributions achieved as a result of galaxy evolution.

We particularly use the following quantities that reflect the evolutionary stage of galaxy: metallicity ($Z$, which include both gas-phase metals and dust), gas mass fraction $f_{\rm gas} \equiv M_{\rm gas}/(M_{\rm gas}+M_{\rm star})$, where $M_\mathrm{gas}$ and $M_\mathrm{star}$ are the masses of gas and stars, respectively, and the specific star formation rate (sSFR = SFR/$M_{\rm star}$).

\subsection{Evolution of grain size distribution}\label{subsec:dustmodel}
We adopt the dust evolution model from \citetalias{Hirashita/Aoyama:2019} and \citetalias{HM20}, whose treatment of grain size distribution was originally developed by \cite{Asano/etal:2013}.
To purely focus on the dust properties predicted by the current galaxy evolution model, we assume that the dust does not affect the other properties of galaxies.
Because of this assumption, it is sufficient to post-process the above semi-analytic calculation results with our dust evolution model. 
From the semi-analytic model, we only know a global quantity for each galaxy; thus, we basically treat each galaxy as a one-zone object.
\citetalias{Huang/etal:2020} already applied this dust model to post-process cosmological simulation results (IllustrisTNG; \citealt{Nelson:2019}), and we follow their post-processing method in this paper.
The fundamental output of this model is the grain size distribution at time $t$, $n(a,\, t)$. 
The grain size distribution is defined such that $n(a,\, t)\,\mathrm{d}a$ is the number of dust grains with radius between $a$ and $a+\mathrm{d}a$ per hydrogen atom.
We simply denote the grain size distribution as $n(a)$ when the time is obvious.
We assume the grains to be spherical and compact, so that the mass of a grain, $m$, is written as $m =(4\pi/3)s a^3$, where $s$ is the material density of the grain.
Since our dust evolution model is not capable of treating multiple grain species, we need to assume dust parameters of a specific species.
Following \citetalias{HM20}, we adopt $s = 2.24\;{\rm g\;cm^{-3}}$ assuming graphite dust \citep{Weingartner/Draine:2001}.
When we calculate the extinction curves, we decompose the grain size distribution into various species by the method we will explain in Section \ref{subsec:ext}.

We calculate the evolution of $n(a,\, t)$ with the initial condition $n(a,\, t=0)=0$, taking into account not only the dust evolution processes (Section \ref{subsubsec:processes}) but also the growth of galaxies (Section \ref{subsubsec:merger}).
We discretize the entire range of grain size ($a = 10^{-4} - 10 \;{\rm \mu m}$) into 32 logarithmically equally separated bins and set $n(a,\, t) = 0$ at the boundaries.

The efficiencies of the dust evolution processes included in our model (Section \ref{subsubsec:processes}), as well as the grain material density, depend on the grain species (mainly through the tensile strength for shattering, grain velocities for shattering and coagulation, and accretion time-scale), although the dependence is not strong.
We adopt graphite properties for the consistency with the choice of $s$ above.
\citetalias{HM20} compared the grain size distributions calculated for silicate and graphite properties, and showed that the resulting grain size distribution is slightly more biased to larger sizes for silicate.
We note, however, that \citetalias{Huang/etal:2020} also adopted the graphite properties for their post-processing, and that some adjustable parameters in grain processing (shattering and coagulation) are degenerate with the adopted grain species.
Therefore, we simply follow \citetalias{Huang/etal:2020} (i.e.\ adopt the graphite properties) for the calculation of grain size distribution for the convenience of comparison with previous results.

The dust-to-gas ratio at time $t$, $\mathcal{D}(t)$, is estimated from the grain size distribution as
\begin{equation}
\mathcal{D}(t) = \frac{1}{\mu m_\mathrm{H}} 
\int {\rm d}a\; \frac{4\pi}{3}a^3 s n(a,\, t),\label{eq:dustgas}
\end{equation}
where $\mu =1.4$ is the gas mass per hydrogen and $m_\mathrm{H}$ is
the mass of hydrogen atom.
The total dust mass, $M_\mathrm{dust}$, is evaluated as $M_\mathrm{dust}=\mathcal{D}M_\mathrm{gas}$.

Below we describe the outline of how to calculate the evolution of grain size distribution.
Since we follow \citetalias{Huang/etal:2020} for the post-processing, we only give a brief summary of the treatment of each process and refer the interested reader to \citetalias{Huang/etal:2020} for the full description of post-processing and to \citetalias{Hirashita/Aoyama:2019} for the full set of basic equations.

\subsubsection{Treatment of merger trees}\label{subsubsec:merger}
Galaxies grow through gas accretion from the circum-galactic medium (CGM) and galaxy merger.
For gas accretion, we assume that the CGM contains no dust and only a few metals. 
Thus, gas accretion changes (dilutes) the dust-to-gas ratio and metallicity.
In a gas accretion episode in which the metallicity changes from $Z(t)$ to $Z(t+\Delta t)$ [$Z(t) > Z(t+\Delta t)$], we update the dust size distribution as
\begin{equation}
n(a,\, t+\Delta t)	= n(a,\, t)\frac{Z(t+\Delta t)}{Z(t)}. 
\end{equation}

If a galaxy merger happens between the two adjacent time-steps $t_1$ and $t_2$, the grain size distribution in the merged object as following.
Suppose that two galaxies with gas masses $M_{\rm gas1}$ and $M_{\rm gas2}$ at $t=t_1$ are merged into a galaxy with gas mass $M_{\rm gas}$ at $t=t_2$.
Assuming that the gas mass ratio of the two galaxies does not change up to a point of merger and that the gas mass is conserved right before and after the merger, we can estimate the gas mass right before the merger, $M_{\rm gas1}^{'}$ and $M_{\rm gas2}^{'}$, from the following two conditions: $M_{\rm gas1}^{'}/M_{\rm gas2}^{'} = M_{\rm gas1}/M_{\rm gas2}$ and $M_{\rm gas1}^{'}+M_{\rm gas2}^{'} = M_{\rm gas}$.
The grain size distributions just before the merger, $n_1^{'}(a)$ and $n_2^{'}(a)$, are calculated from the grain size distributions in the previous snapshot (at $t=t_1$), $n_1(a)$ and $n_2(a)$, using our dust evolution model (Section \ref{subsubsec:processes}).
Finally we estimate the grain size distribution after the galaxy merger as
\begin{equation}
n(a) = \frac{M_{\rm gas1}^{'}n_1^{'}(a)+M_{\rm gas2}^{'}n_2^{'}(a)}{M_{\rm gas}}.
\end{equation}

\subsubsection{Dust evolution processes}\label{subsubsec:processes}
Now we explain the outline of how to calculate the evolution of $n(a,\, t)$ for each galaxy.
We consider five processes for dust evolution: stellar dust production (dust condensation in stellar ejecta), dust destruction by SN shocks, grain growth by coagulation (grain--grain sticking), dust growth by the accretion of gas-phase metals, and grain disruption by shattering. 
The increment of dust abundance by stellar dust production is calculated based on the metallicity increase by assuming a constant condensation efficiency $f_\mathrm{in}=0.1$. 
The grain size distribution of this newly formed dust is assumed to be a lognormal function with an average grain radius of 0.1 $\micron$ and a standard deviation of 0.47 (in the log). 
The decrement of the grain size distribution by SN dust destruction is estimated based on a grain-size-dependent destruction efficiency multiplied by the rate at which the interstellar dust is swept up by SN shocks\citep{McKee:1989}. The destruction rate proportional to the SN shock sweeping time is often adopted in one-zone models \citep[including post-processing of merger trees; e.g.][]{Lisenfeld:1998,Dwek:1998,Huang/etal:2020} as well as in hydrodynamic simulations \citep[e.g.][]{Aoyama/etal:2017}.

Each of the other three processes (coagulation, accretion, and shattering) occurs in a specific ISM phase. 
We assume that the ISM consists of two phases, namely diffuse and dense components with $(n_{\rm H}/{\rm cm}^{-3},\, T_{\rm gas}/{\rm K})=(0.3,\, 10^4)$ and $(300,\, 25)$, respectively, where $n_{\rm H}$ is the hydrogen number density, and $T_{\rm gas}$ is the gas temperature.
Since $\nu^2$GC model does not distinguish between the dense and diffuse phases, we need to evaluate the dense gas fraction $\eta_{\rm dense}$, which we assume to be related to the star formation efficiency (\citetalias{Huang/etal:2020}):
\begin{equation}
\eta_{\rm dense} = \tau_{\rm ff}\frac{\mathrm{SFR}}{\epsilon_{*}M_{\rm gas}},\label{eq:dense}
\end{equation}
where $\tau_{\rm ff} = 2.51 (n_{\rm H}/300\;{\rm cm}^{-3})^{-1/2}$~Myr is the free-fall time in the dense ISM, and $\epsilon_{*}$ is the star-formation efficiency.
We fix $n_{\rm H} = 300\;\mathrm{cm}^{-3}$ (the density adopted for the dense phase) and $\epsilon_* = 0.01$.
We assume that coagulation and accretion occur in the dense phase while shattering takes place in the diffuse phase.
This is realized numerically by dividing a single time-step $\Delta t$ into $\eta_\mathrm{dense}\Delta t$ for coagulation and accretion and $(1-\eta_\mathrm{dense})\Delta t$ for shattering.
Following \citetalias{Huang/etal:2020}, we set $\omega_{\rm coag} = 10$, where $\omega_{\rm coag}$ is a parameter to control the efficiency of coagulation. \citetalias{Huang/etal:2020} found that this value of $\omega_{\rm coag}$ best describes the extinction curve of MW-like galaxies.

\subsection{Extinction curve}\label{subsec:ext}
We calculate the extinction curve as a representative observable quantity that reflects the grain size distribution.
As noted above, the grain size distribution $n(a)$ is the total of all the grain species. 
In order to calculate the extinction curve, it is crucial to specify the grain species. 
The extinction per hydrogen at wavelength $\lambda$, denoted as $A_\lambda$,\footnote{Usually, $A_\lambda$ denotes the total extinction in a line of sight; thus, it should be multiplied by the hydrogen column density.
However, since we are only interested in the wavelength dependence of extinction we always show the extinction normalized to the value in the $V$ band, $A_\lambda /A_V$. In this case, the hydrogen column density is always cancelled out.} is calculated as
\begin{equation} \label{eq:extinction}
    A_\lambda = 2.5 \log_{10} \mathrm{e} \sum_i \int^\infty_0 n_i(a) \pi a^2 Q_\mathrm{ext}^{(i)}(a,\lambda)\;\mathrm{d}a,
\end{equation}
where $i$ indicates the different grain species and $Q_\mathrm{ext}^{(i)}(a,\lambda)$ is the extinction efficiency factor of the $i$th species evaluated with Mie theory \citep{Bohren:1983}. 
For the grain species, we consider silicate and carbonaceous dust. For the former species,
we adopt astronomical silicate taken from \citet{Weingartner/Draine:2001} (originally \citealt{Draine:1984,Laor:1993}).
The carbonaceous dust is divided into two species: graphite and amorphous carbon, for which we adopt the optical properties from \citet{Weingartner/Draine:2001} and \citet[][ACAR]{Zubko:1996}, respectively. Graphite has a strong feature at 2175 \AA. The decomposition into these species is described in what follows.

The silicate mass fraction is described by (Appendix \ref{app:fsil})
\begin{align}
    f_\mathrm{sil}=0.79+0.13\tanh\left[\log_{10}\left(Z^{-1}\,\frac{\mathrm{sSFR}}{\mathrm{yr}^{-1}}\right)+6.8\right] .\label{eq:fsil}
\end{align}
This fitting formula is derived based on the results presented by \citetalias{HM20}, who evaluated the silicate fraction from the elemental abundance ratio between Si and C.
Note that we use the absolute metallicity (not normalized to solar) for $Z$ and express sSFR in units of yr$^{-1}$.
Carbonaceous dust is further divided into graphite and amorphous carbon with the fraction of graphite, $f_\mathrm{gr}$, as (\citetalias{HM20,Huang/etal:2020})
\begin{align}
    f_\mathrm{gr}=1-\eta_\mathrm{dense}.
\end{align}
As shown by \citetalias{HM20} (based on \citealt{Murga:2019}), this equation approximately holds because the formation of carbonaceous dust with a regular atomic structure (here it is graphite) predominantly occurs in the diffuse ISM.

\section{Basic comparisons with observations}
\label{sec:results_dust}

In this section, we focus on the quantities that can be compared with observations. 
At $z\sim 0$, some detailed comparisons with extinction curves are possible. 
We compare the calculated extinction curves with the observations in the MW, the Large Magellanic Cloud (LMC), and the Small Magellanic Cloud (SMC), whose extinction curves have often been used for comparison with dust models \citep[e.g.][]{Pei:1992,Weingartner/Draine:2001}. 
Although other tests for the grain size distribution are difficult, we still examine the total dust abundance (or mass) using the following two statistical properties: the relation between dust-to-gas ratio and metallicity at $z=0$ and the dust mass functions at various redshifts. 
These comparisons will give a basis on which we make predictions for wider ranges of galaxy parameters (metallicity, redshift, etc.) in the next section.

In the following, we limit our galaxy sample to star-forming galaxies, which are defined to have ${\rm sSFR} > 10^{-11}\;{\rm yr}^{-1}$ and $M_{\rm star} > 10^8\;M_{\odot}$, unless otherwise noted.
Quenched galaxies, whose sSFRs are lower than the above range, are dust-poor with negligible contribution to the total dust budget in the Universe. Moreover, for such quenched galaxies, dust evolution is not necessarily linked to the star formation and metal enrichment \citep{Hirashita:2015_ell,Hirashita:2017}.
For these reasons, we neglect quenched galaxies, and only focus on star-forming galaxies.

\subsection{Extinction curves in nearby galaxies}\label{subsec:ext_nearby}

\begin{figure*}
  \includegraphics[width=2\columnwidth]{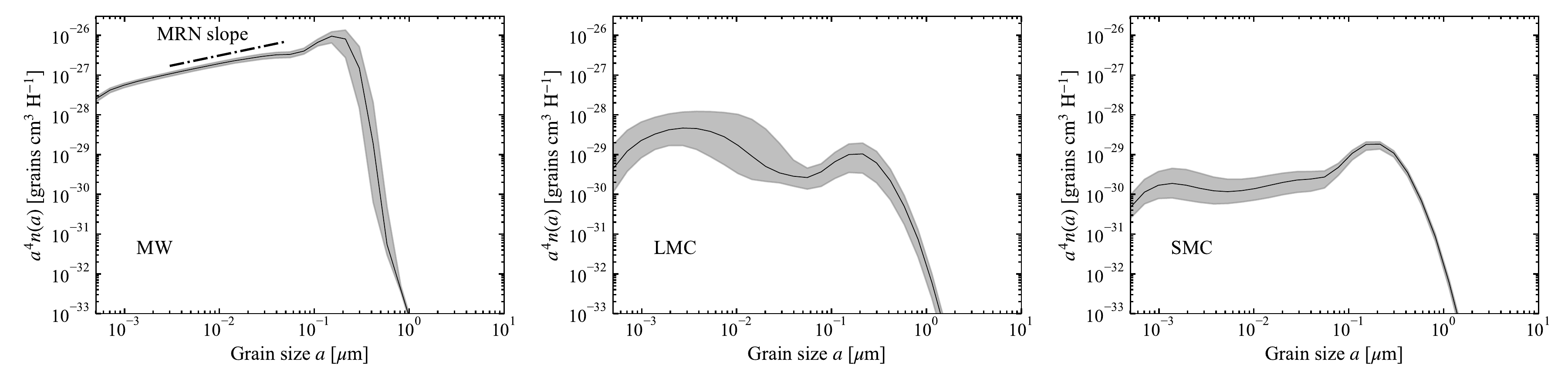}
  \includegraphics[width=2\columnwidth]{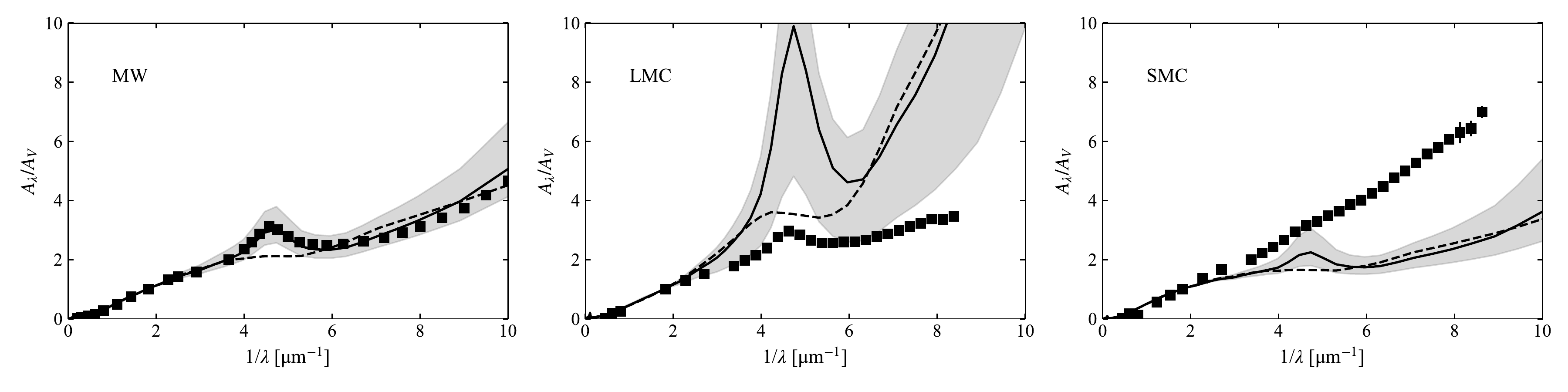}
  \caption{Upper panels: Grain size distribution of MW-like galaxies (left), LMC-like galaxies (middle) and SMC-like galaxies (right) in the mock.
  The grain size distribution (per hydrogen) is multiplied by $a^4$, so that the resulting quantity is proportional to the grain mass abundance per $\log a$.
  In each panel, the solid line shows the median, while the shaded region presents the 25th and 75th percentile range.
  The dot--dashed line in the left panel shows the slope of the \citetalias{MRN} grain size distribution [$n(a)\propto a^{-3.5}$].
  Lower panels: Extinction curves of the MW (left), LMC (middle) and SMC (right) analogues in the mock.
  In each panel, the solid line shows the median, while the shaded region presents the 25th and 75th percentile range.
  The black squares show observational data taken from \protect\cite{Pei:1992} for the MW and \protect \cite{Gordon/etal:2003} for the LMC and SMC.
  The dashed lines show the median extinction curves in the case where we use amorphous carbon for all the carbonaceous dust.
  \label{fig:size_MW}}
\end{figure*}

First we examine the MW-like galaxies to validate the model.
We define the MW analogues in our generated sample at $z=0$ by similar stellar mass and sSFR to those of the MW.
We adopt $M_{\rm star} = (6.08 \pm 1.14) \times 10^{10} M_\odot$ and sSFR $= (2.71 \pm 0.59) \times 10^{-11} {\rm yr^{-1}}$ for the value of the MW \citep{Licquia/Newman:2015}.
We sample galaxies within the 3$\sigma$ ranges of these values from the $\nu^2$GC mock catalog at $z=0$ and obtain 558 MW-like galaxies.

The upper left panel of Fig.\ \ref{fig:size_MW} shows the grain size distribution of the MW-like galaxies in the mock.
The grain sizes distributions of the MW analogs are consistent with the \citetalias{MRN} distribution $n(a) \propto a^{-3.5}$, which is inferred from the observed extinction curve of the MW.
Moreover, the upper cut-off of the grain radius, $a=0.25~\micron$, in the \citetalias{MRN} distribution is also reproduced well in our model.
The dispersion is small, which means not only that the overall dust abundance level is similar in the MW-like sample, but also that the shape of the grain size distribution is robustly determined. 
This robust convergence to the \citetalias{MRN} shape is consistent with other calculations (e.g.\ \citetalias{Hirashita/Aoyama:2019}), in which the balance between coagulation and shattering leads to an \citetalias{MRN}-like grain size distribution.

In the lower left panel of Fig.\ \ref{fig:size_MW}, we show the extinction curves of the MW-like sample and compare them with the observed MW curve taken from \citet{Pei:1992}. 
The calculated extinction curves are in good agreement with the observation, which is attributed to the above-mentioned robust prediction of the \citetalias{MRN} grain size distribution.
Note that the fractions of various grain compositions are predicted quantities in our model (Section \ref{subsec:ext}). 
Thus, the good match of the bump at 2175 \AA\ (formed by graphite) and the UV slope (contributed from small grains of all species) of the MW extinction curve indicates that the fractions of various grain components are also validated.

Based on the above success, we also extend the comparison to the LMC and SMC.
We select analogues of the LMC and SMC in the mock by stellar mass and sSFR.
We adopt $\log (M_{\rm star}/M_\odot) = 9.3 \pm 0.1 $ and $\log ({\rm sSFR/yr^{-1}}) = -9.7 \pm 0.22$ for the LMC and $\log (M_{\rm star}/M_\odot) = 8.5 \pm 0.1 $ and $\log ({\rm sSFR/yr^{-1}}) = -10.1 \pm 0.22$ for the SMC \citep{Skibba/etal:2012}.
We sample galaxies within 3$\sigma$ ranges of these values from the mock catalog at $z = 0$ and obtain 1782 LMC-like galaxies and 4332 SMC-like galaxies.
We find that the slope and the 2175 \AA\ bump strength of the model extinction curves do not match to the observations.
The overproduction of the bump strength could be due to the uncertainty in the graphite fraction in our model. 
Thus, we also show the extinction curve with $f_{\rm gr} = 0$; i.e.\ all the carbonaceous dust is amorphous, as an extreme case.
In this case the bump is suppressed as expected, but the observed slope of extinction curves are not reproduced.

In fact, the observed extinction curves of the LMC and SMC lie between the two predictions in our model. 
Therefore, the mismatch may be due to the rapid evolution from flat to steep extinction curves, which is driven by the fast increase of small grains by accretion (dust growth; see \citetalias{Hirashita/Aoyama:2019}). 
The rapid evolution may be partly due to the one-zone treatment: if we consider the spatial inhomogeneity in dust evolution within a galaxy, the increase of small grains does not occur coherently, so that we expect that the evolution from flat to steep extinction curves could become moderate. 
Indeed, as shown by \citet{Aoyama:2020}, the grain size distributions are different between the dense and diffuse ISM phases, indicating that the coherent evolution is not likely to occur in reality. 
This may be true even if we include the effect of diffusion, which causes mass exchange between the two phases \citep{Romano:2022_diffusion}. 
Therefore, we defer the solution of the mismatch to the LMC and SMC extinction curves to future spatially resolved simulations.
Alternatively, physical processes not included in our calculations may be responsible for the discrepancy:
Different radiation force strengths between silicate and carbonaceous dust may lead to the selective loss of small carbonaceous grains, which could explain the weak 2175 \AA\ bump strengths in the LMC and SMC \citep{Bekki:2015}.

We should also keep in mind that the LMC and SMC are satellite galaxies. 
The LMC and SMC also interact with each other \citep[e.g.][]{Fujimoto:1977}.
Thus, the mismatch could be due to the insufficient modeling of the environment effect on satellite galaxies.
The gravitational interaction could expel gas and dust from the satellite galaxies and could also affect their star formation properties. 
This would directly affect our selection criteria for the LMC- and SMC-like galaxies. 
Future cosmological simulations that naturally treat such interactions could serve to further investigate the dust properties in the LMC and SMC.

\subsection{Dust abundance}\label{subsec:abundance}

\begin{figure}
  \includegraphics[width=\columnwidth]{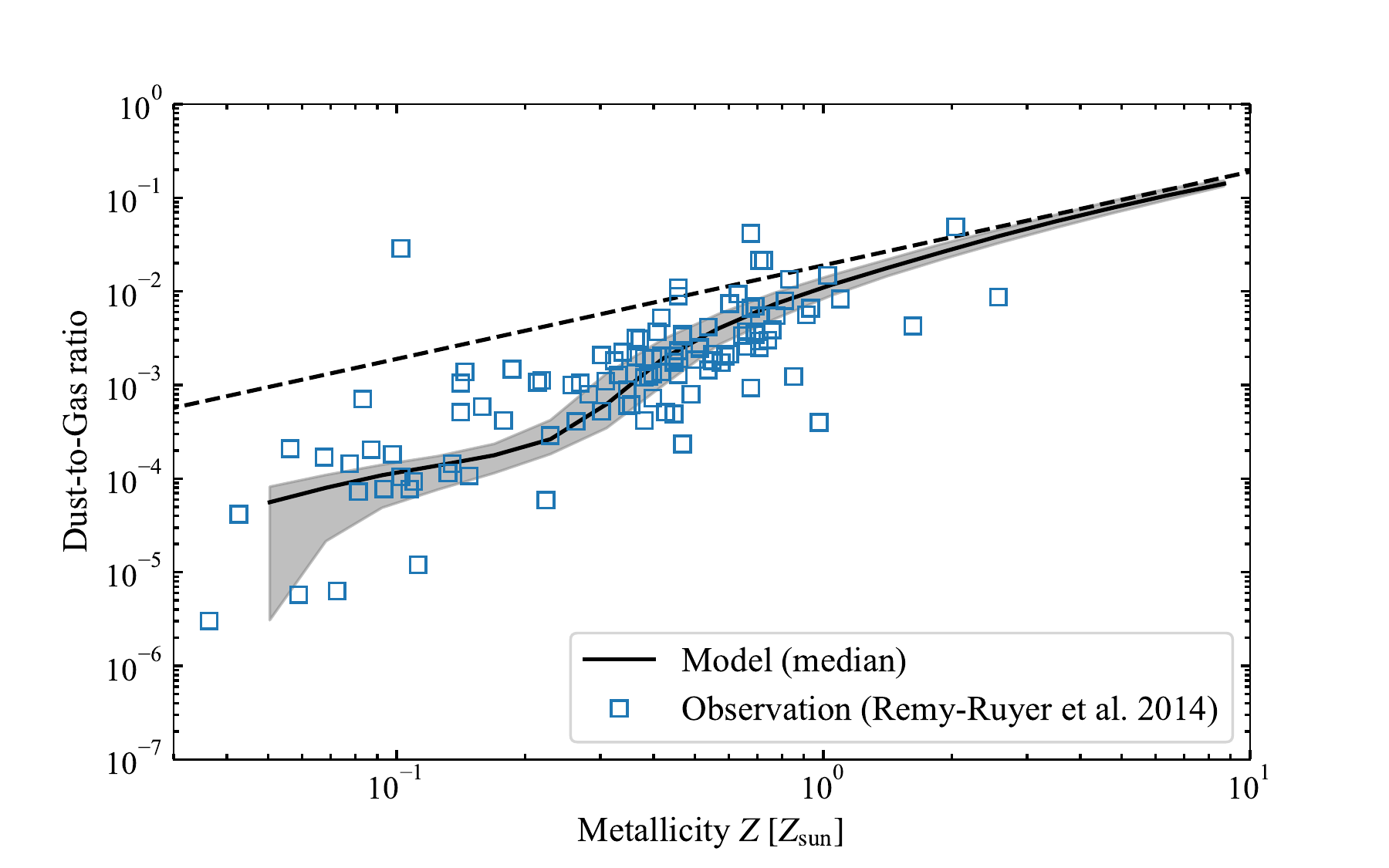}
  \includegraphics[width=\columnwidth]{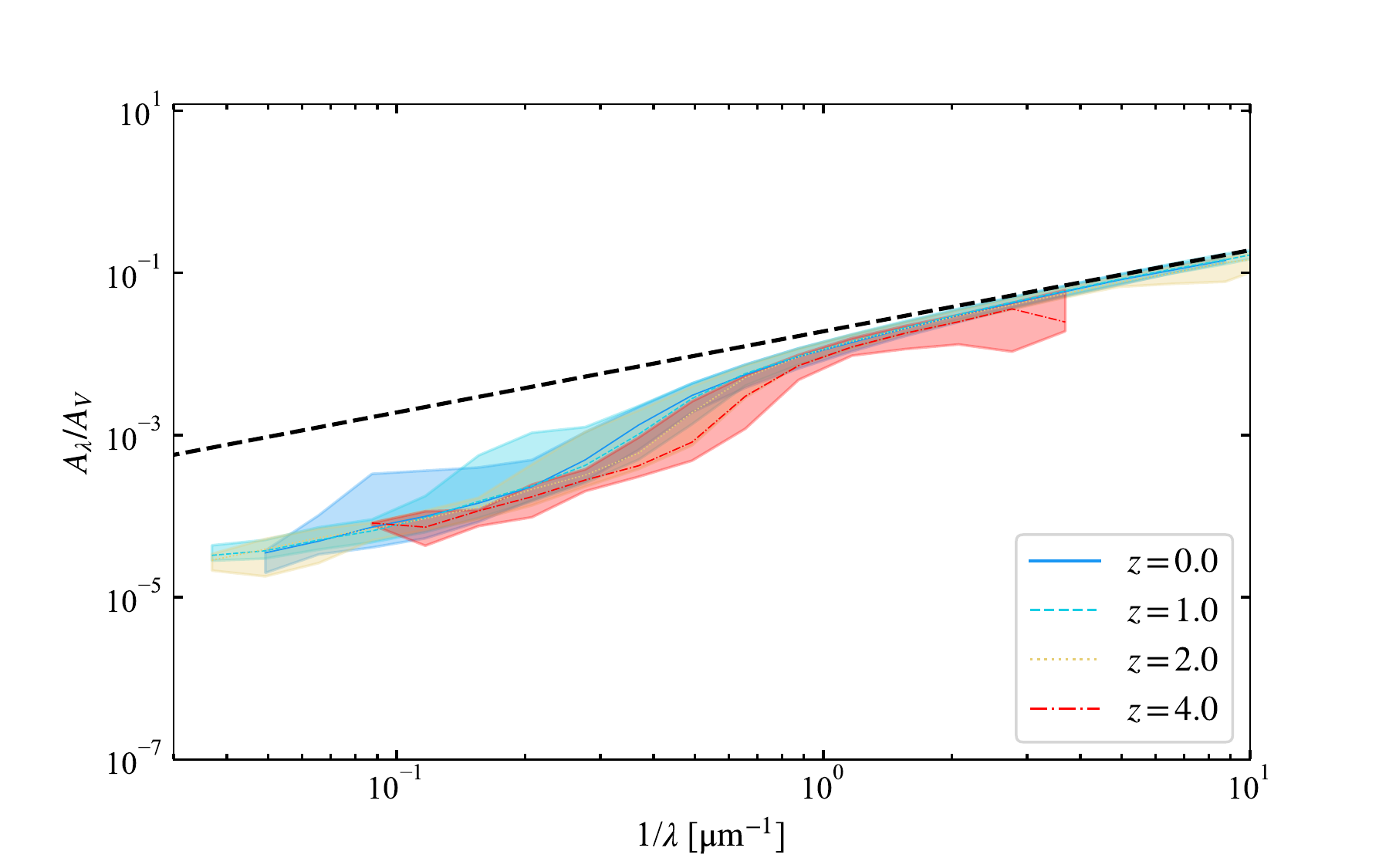}
  \caption{Upper: Relation between dust-to-gas ratio and metallicity.
	The solid line shows the median, and the shaded region presents the 5th and 95th percentile range (because of the small dispersion, we take a wide percentile range) of the dust-to-gas ratio at each value of the metallicity. The dashed line shows the upper limit of the dust-to-gas ratio ($\mathcal{D}=Z$). The blue open squares are the observational data taken from \protect\cite{RemyRuyer/etal:2014}.
	Lower: Redshift evolution of the relation between dust-to-gas ratio and metallicity from $z=4$ to $z=0$.
  The shaded regions show the 25th and 75th percentile range. Each colour corresponds to a redshift as shown in the legend.\label{fig:DZ}}
\end{figure}

We also test the total dust abundance represented by the dust-to-gas ratio (equation \ref{eq:dustgas}) and the total dust mass. 
In particular, the relation between dust-to-gas ratio and metallicity is often examined since dust evolution is strongly related to the metal enrichment \citep[e.g.][]{Lisenfeld:1998,Dwek:1998}.
The upper panel of Fig.\ \ref{fig:DZ} shows the relation between dust-to-gas ratio and metallicity at $z=0$.
The dust-to-gas ratio increases with metallicity and approaches its upper limit, $\mathcal{D} = Z$. 
The steep increase of dust-to-gas ratio at sub-solar metallicity is due to dust growth by accretion.
The overall trend of the model galaxies are consistent with the observed relation obtained by \cite{RemyRuyer/etal:2014}.
Recent analysis of nearby galaxy data also obtained a similar relation \citep{Galliano:2022}.
Thus, our model reproduces the coevolution of dust and metals well.

As a natural extension of the above result,
we show the redshift evolution of the relation between dust-to-gas ratio and metallicity in the lower panel of Fig.\ \ref{fig:DZ}. 
There is a slight trend that, at a fixed metallicity, the dust-to-gas ratio is smaller at $z=4$ than at lower redshifts as also predicted in other models \citep[e.g.][]{Vijayan/etal:2019,Triani:2020}. 
However, the $\mathcal{D}$--$Z$ relation is broadly unchanged at $z\lesssim 2$.
The lower dust-to-gas ratio at $z\sim 4$ is probably due to the shorter star formation time-scale:
as shown by \citet{Asano:2013a}, if the star formation occurs quickly, the dust has less time for interstellar processing, especially dust growth by accretion, leading to lower dust abundance.
The redshift evolution of the $\mathcal{D}$--$Z$ relation is not as clear as that shown in a cosmological simulation by \citet{Hou:2019}. 
Our result may be rather consistent with the semi-analytic model by \citet{Popping/etal:2017}, who showed little evolution of the $\mathcal{D}$--$Z$ relation with redshift.
Observations of quasar absorption lines also indicated little evolution from $z\sim 0$ to $z\sim 5$ \citep{Peroux:2020,Popping:2022a}.
These unchanged $\mathcal{D}$--$Z$ relations may result from the robust balance between dust production and destruction \citep[e.g.][]{Inoue:2011,Mattsson:2014}.
The difference between the cosmological simulation and the semi-analytic models could be interpreted as the importance of local (interstellar) physical conditions, which are not included in the semi-analytic treatment.

\begin{figure*}
  \includegraphics[width=2\columnwidth]{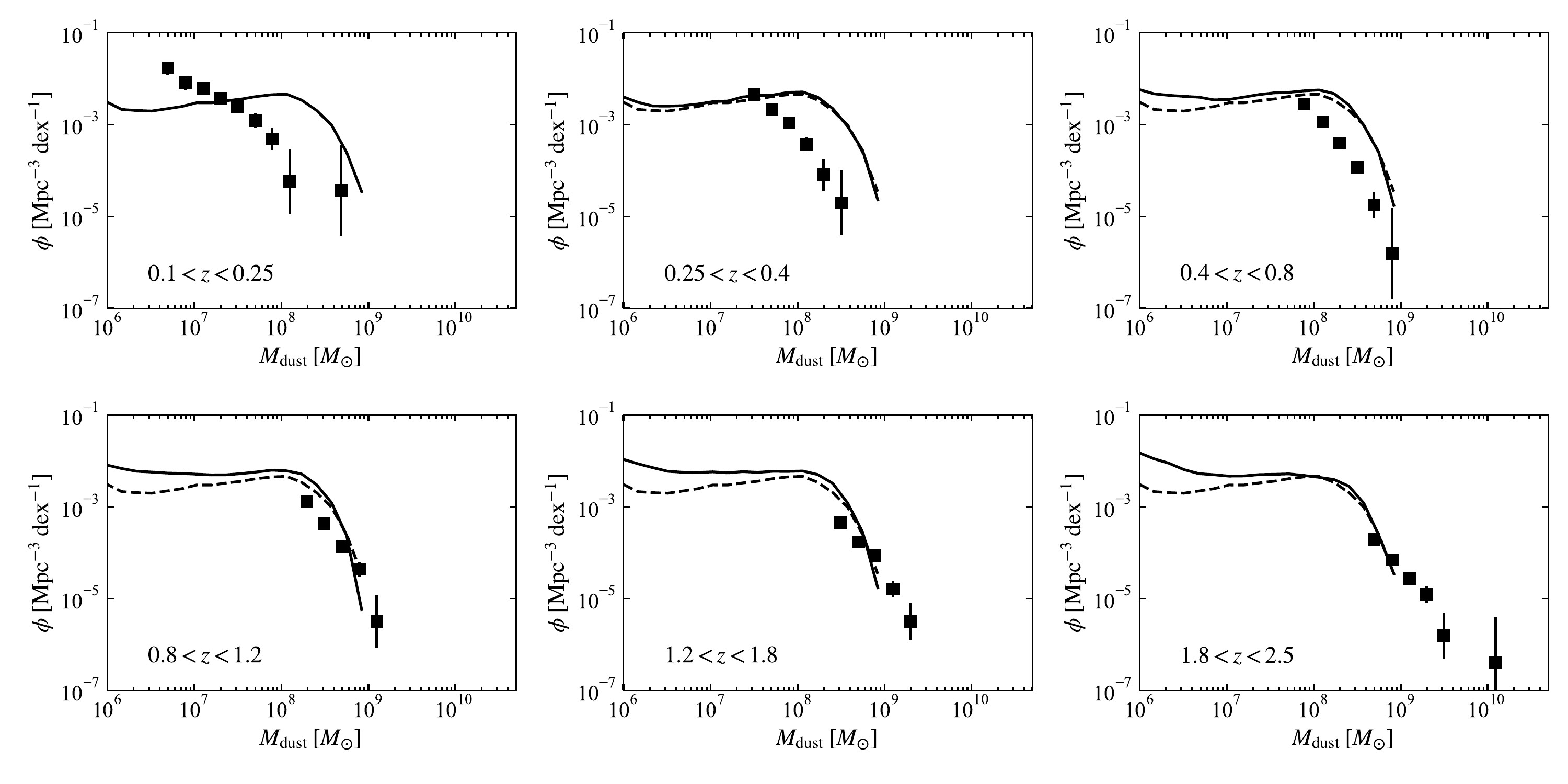}
  \caption{\label{fig:dust_MF_z}
  Dust mass functions at various redshifts.
  In each panel, the solid line shows the model prediction, which is compared with the measurements (squares) obtained by
  \protect\cite{Pozzi/etal:2020}.
  For our results, we use the snapshots at $z = 0.19$, 0.36, 0.61, 1.0, 1.5, and 2.1. The redshift range of the observational data is indicated in each panel.
  We also show the model prediction at $z = 0.19$ in each panel by the dashed line in order to clarify the redshift evolution.
  }
\end{figure*}

We also test the statistical properties of dust mass, specifically the dust mass function (distribution function of dust mass) at various redshifts in Fig.\ \ref{fig:dust_MF_z}. 
The observed dust mass functions for comparison are taken from \cite{Pozzi/etal:2020}, who estimated the dust mass of a galaxy sample in the COSMOS field using SED fitting to the \textit{Herschel} data in a redshift range of $z=0$--2.5.
We observe in Fig.\ \ref{fig:dust_MF_z} that the model overpredicts the number density of dust-rich galaxies at $z < 0.8$ while it roughly reproduces the massive end of the observed dust mass functions at higher redshifts.
Our mock catalog does not contain dusty galaxies with $M_\mathrm{dust}>10^9$ M$_{\sun}$ at $z>1.2$ because such massive galaxies are too rare to be included in our simulation box (recall that our semi-analytic model is based on an $N$-body simulation; Section \ref{subsec:nu2GC}).

The overprediction of dust mass function at high $M_\mathrm{dust}$ in the present Universe is also seen in cosmological simulations by \citet{Aoyama/etal:2018} and \citet{Hou:2019}. According to \citet{Hou:2019}, the overprediction is due to
the excessive amount of gas retained in massive galaxies.
However, we confirmed that the gas mass function is reproduced well in our semi-analytic model \citep{Makiya/etal:2016}.
Thus, we suspect that the overprediction of high-dust-mass galaxies is due to the overprediction of the dust-to-gas ratio.
Indeed, $\nu^2$GC overpredicts the cold gas metallicity of the high stellar mass galaxies ($M_{\rm star} \gtrsim 10^{11} {\rm M}_{\odot}$) about factor of 2 at $z = 0$ (fig.\ 17 of \citealt{Makiya/etal:2016}).
This could lead to overestimate the dust masses of massive galaxies. 
The overproduction of metals can also be seen in the extended $\mathcal{D}$--$Z$ relation beyond the high-metallicity end of the observational data in Fig.\ \ref{fig:DZ}.
This overestimating problem of metallicity may be related to the difficulty in reproducing the star formation histories of individual galaxies; for example, $\nu^2$GC underestimates the specific SFRs of individual galaxies at $z\sim 2$ in a way similar to the cosmological simulation performed by \citet{Bassini:2020}.
So far, no model has reproduced the dust mass functions at $z\sim 0$ and at $z\gtrsim 1$ simultaneously. 
In general, there is a tendency that a model overproduces the dust mass in massive galaxies at $z\sim 0$ while it underpredicts massive dusty galaxies at $z\gtrsim 1$; or that, if the model reproduce the dust mass function at low/high redshift, it fails at high/low redshift \citep{McKinnon:2017,Popping/etal:2017,Triani:2020}.

\begin{figure}
  \includegraphics[width=\columnwidth]{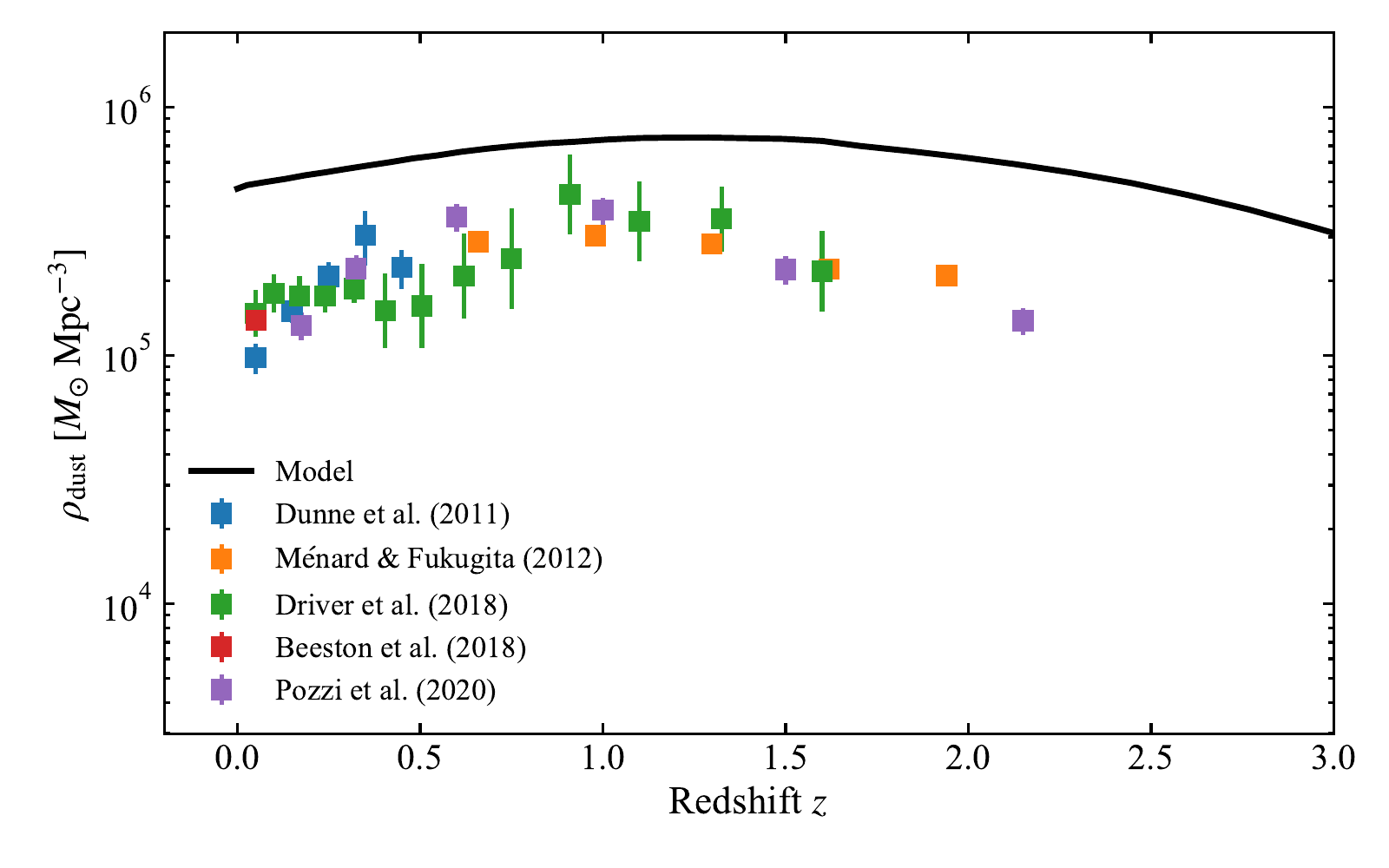}
  \caption{
  Comoving dust mass density in the Universe as a function of redshift. Data points are the measurements obtained by \protect\cite{Dunne/etal:2011}, \protect\cite{Menard/Fukugita:2012}, \protect\cite{Driver/etal:2018}, \protect\cite{Beeston/etal:2018} and \protect\cite{Pozzi/etal:2020} as indicated in the legend while the black solid line shows our model prediction.\label{fig:rho_dust}}
\end{figure}

For the statistical properties of dust mass, we also examine the comoving dust mass density in the Universe in Fig.\ \ref{fig:rho_dust}.
For the comoving dust mass density, we sum up all the dust mass contained in the galaxies and divide it by the total comoving volume. 
We also collect observational data in the literature that basically derived the comoving dust mass density in a similar way; that is, by estimating the dust mass contributed from all the galaxies at each redshift \citep{Dunne/etal:2011,Driver/etal:2018,Beeston/etal:2018,Pozzi/etal:2020}.
In addition, we plot dust mass contained in Mg \textsc{ii} absorbers \citep{Menard/Fukugita:2012}.
We observe that both our model and the observations show a peak at $z\sim 1$--1.5. 
Thus, the qualitative behaviour of the cosmic dust mass evolution is reproduced well.
Fig.\ \ref{fig:rho_dust} shows, however, that our model overpredicts the comoving dust mass density.
At low redshift, this overprediction is attributed to the overestimated dust mass in massive galaxies in our model (Fig.\ \ref{fig:dust_MF_z}). 
At high redshift ($z\gtrsim 1$), the discrepancy is likely due to the contribution from low-mass galaxies since we reproduce the dust mass functions at high $M_\mathrm{dust}$. However, the dust masses in low-mass galaxies are difficult to obtain observationally, which implies that observational data have a significant uncertainty at $z\gtrsim 1$. 
Moreover, the discrepancy is only a factor of $\sim 2$, which is also comparable to the uncertainty in the dust mass absorption coefficient \citep{Hirashita:2014}. 
Therefore, we conclude here that our model reproduces the observed total dust mass in the Universe within a factor of $\sim 2$.

From the above, we argue that our model is as good as other theoretical models in terms of reproducing the statistical properties of the dust mass.
Some issues possibly related to overprediction of metallicity at the massive end and to observational uncertainties should be addressed in the future studies. 
Keeping these issues in mind, we return to the grain size distribution, which is the central quantity in our model, and further examine its evolution and its relation to galaxy properties.

\section{Predicted statistical properties}\label{sec:prediction}

In this section, we provide the predictions from our model. We concentrate on how the grain size distribution, which we focus on in this paper, is related to galaxy properties. 
We also predict the extinction curve as an observable property that reflect the grain size distribution.

\subsection{Grain size distribution}\label{subsec:gsd}

\begin{figure*}
  \includegraphics[width=2\columnwidth]{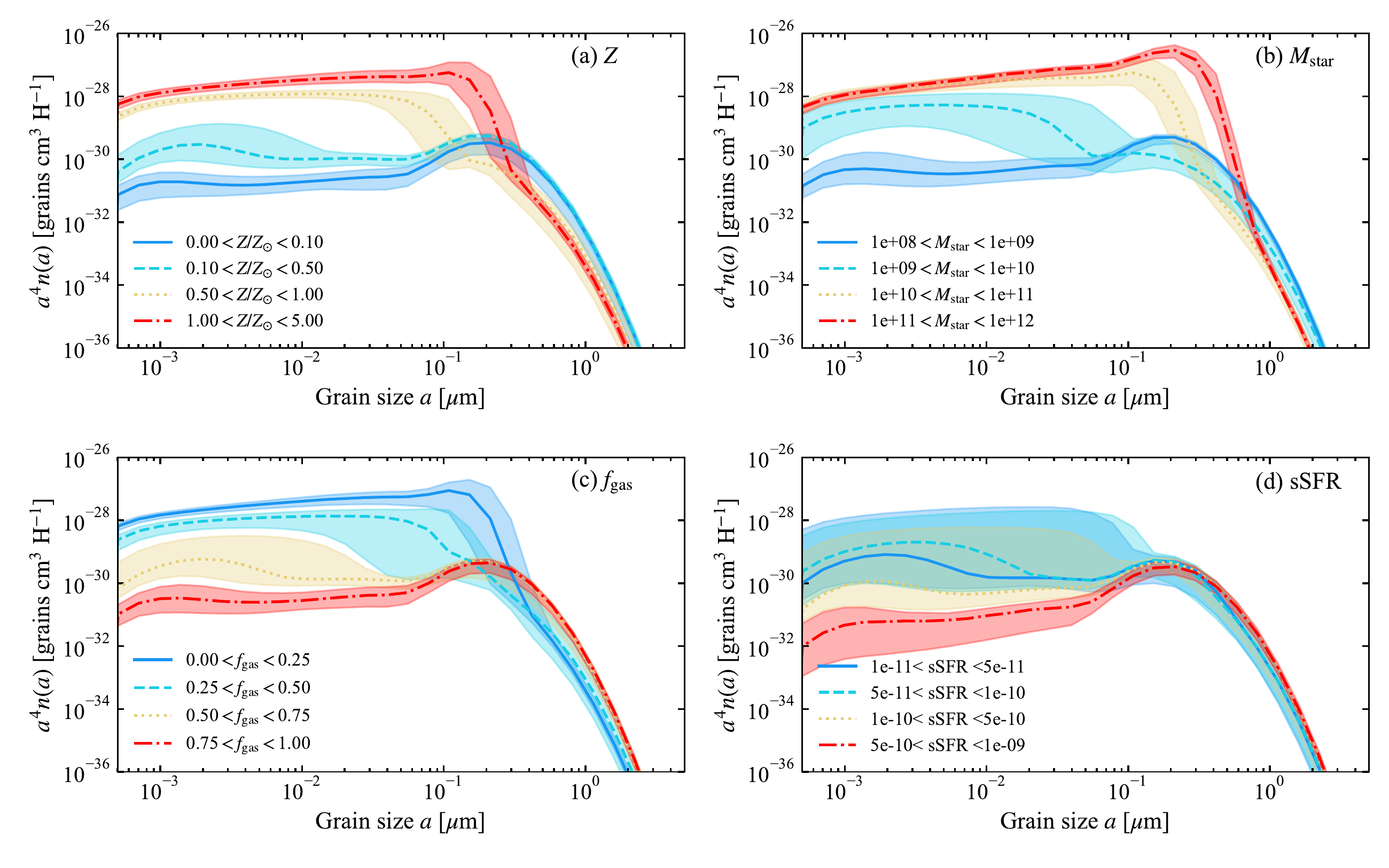}
  \caption{\label{fig:size_dependence}
  Grain size distributions of galaxies binned by the following quantities: metallicity (top left), stellar mass (top right), gas fraction (bottom left), and sSFR (bottom right). Each colour presents each bin as shown in the legend.
  The lines show the medians, while the shade regions correspond to the 25th and 75th percentile ranges.}
\end{figure*}

We examine the relation between the grain size distribution and galaxy properties, for which we consider the metallicity, stellar mass, gas fraction, and sSFR, in order to examine how the grain size distribution evolves along with the build-up of galaxies.
To this goal, we show the grain size distributions in four bins for each quantity at $z=0$ in Fig.\ \ref{fig:size_dependence}.
In general, as a galaxy evolves, the stellar mass and metallicity increase while the gas fraction and sSFR decrease.
The figure clearly shows strong relations between the grain size distribution and each of these four quantities. As shown by previous studies (\citealt{Asano/etal:2013}; \citetalias{Hirashita/Aoyama:2019,HM20}), the grain size distribution is dominated by large grains (the peak at $a\sim 0.1~\micron$ formed by stellar dust production is clear) in the early stage of the evolution, the abundance of small grains increases by accretion in the intermediate stage, and small grains are converted to large grains by coagulation in the late stage. 
In the end, the grain size distribution approaches an \citetalias{MRN}-like shape as also observed for the MW-like galaxies in Section \ref{subsec:ext_nearby}.
Overall, the grain size distribution depends strongly on the global parameters (i.e.\ the above four quantities), and this dependence overwhelms the dispersion in the grain size distribution among the sample. 
This means that, if we measure one of those four quantities, we are able to predict the grain size distribution well.

Among the four quantities, sSFR has the loosest relation with the grain size distribution. This is because sSFR has a poor correlation with $Z$. Many star-forming galaxies at various evolutionary stages lie on the so-called main sequence with a constant value of sSFR. 
Thus, sSFR is less useful than the other three quantities in characterizing the grain size distribution.

\begin{figure*}
  \includegraphics[width=2\columnwidth]{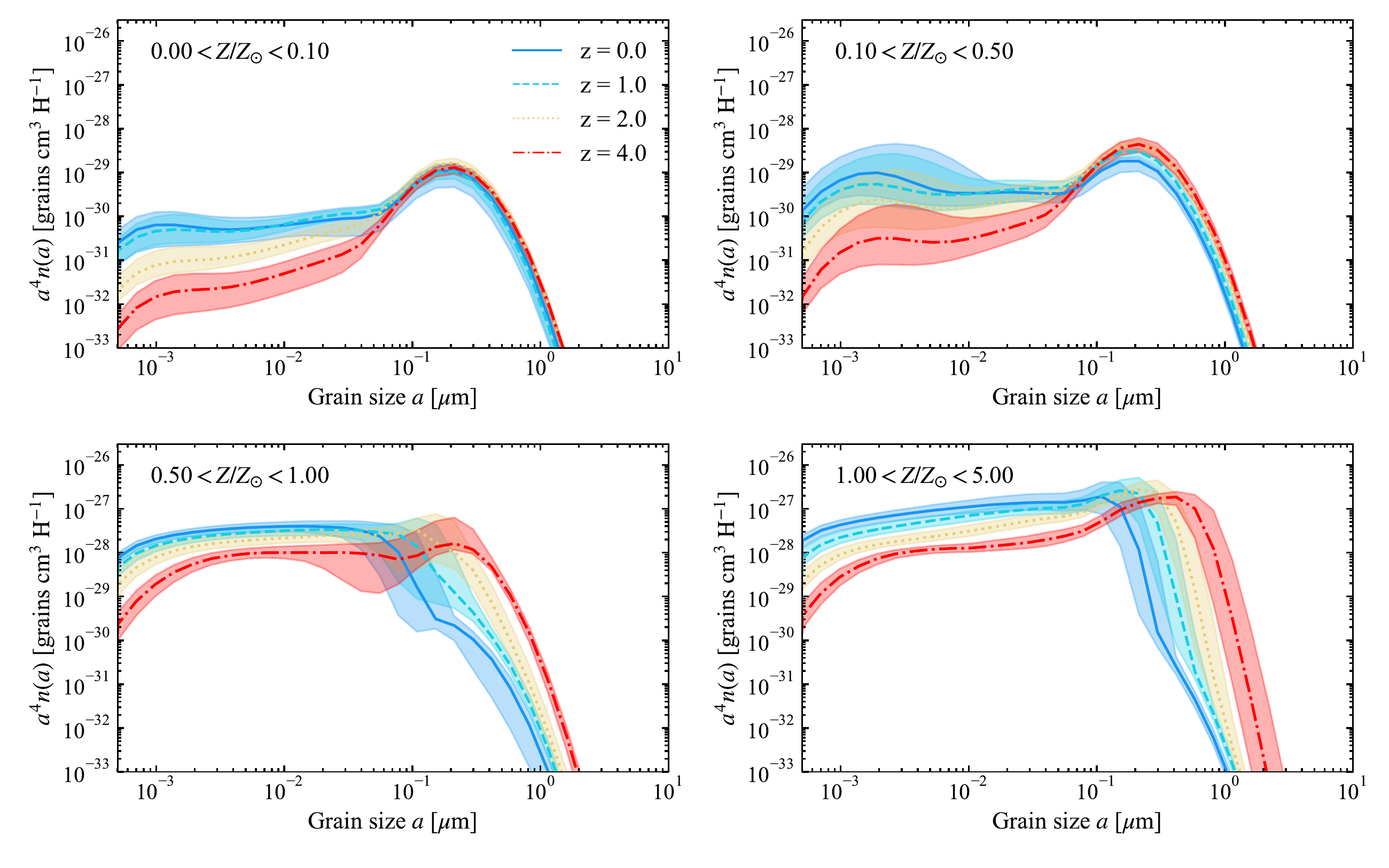}
  \caption{\label{fig:size_z}
  Redshift evolution of grain size distribution binned by the metallicity (shown in the upper left corner in each panel).
  The lines show the median in each bin, while the shaded regions present the 25th and 75th percentile range.
  Each colour correspond to each redshift as shown in the legend.}
\end{figure*}

We examine the redshift evolution of grain size distribution in different bins of the global parameters. 
Since all these parameters are correlated, we only show the metallicity bins, and the redshift trends for the other parameters are similar to that for the metallicity.
Fig.\ \ref{fig:size_z} shows the redshift evolution of grain size distribution binned by the metallicity at $z = 0.0$.
We find that low-redshift galaxies tend to have more small grains compared to high-redshift galaxies with similar metallicity.
At $Z>0.5$ Z$_{\sun}$, the upper cut-off of the grain size distribution lies at a smaller grain radius at lower redshift because of shattering. Thus, the abundance ratio between small and large grains increases monotonically with redshift at any metallicity.
This trend is also seen in the small-to-large grain abundance ratio in a cosmological simulation by \citet{Hou:2019}.
We also note that at any redshift there is a clear trend with metallicity from large-grain-dominated to \citetalias{MRN}-like shapes.
We also observe that the metallicity dependence of grain size distribution converges at $z\lesssim 1$. 
Thus, the grain size distribution strongly evolves at $z>1$, and the present-day relation between grain size distribution and metallicity emerges at $z<1$.

\subsection{Extinction curves}

\begin{figure*}
  \includegraphics[width=2\columnwidth]{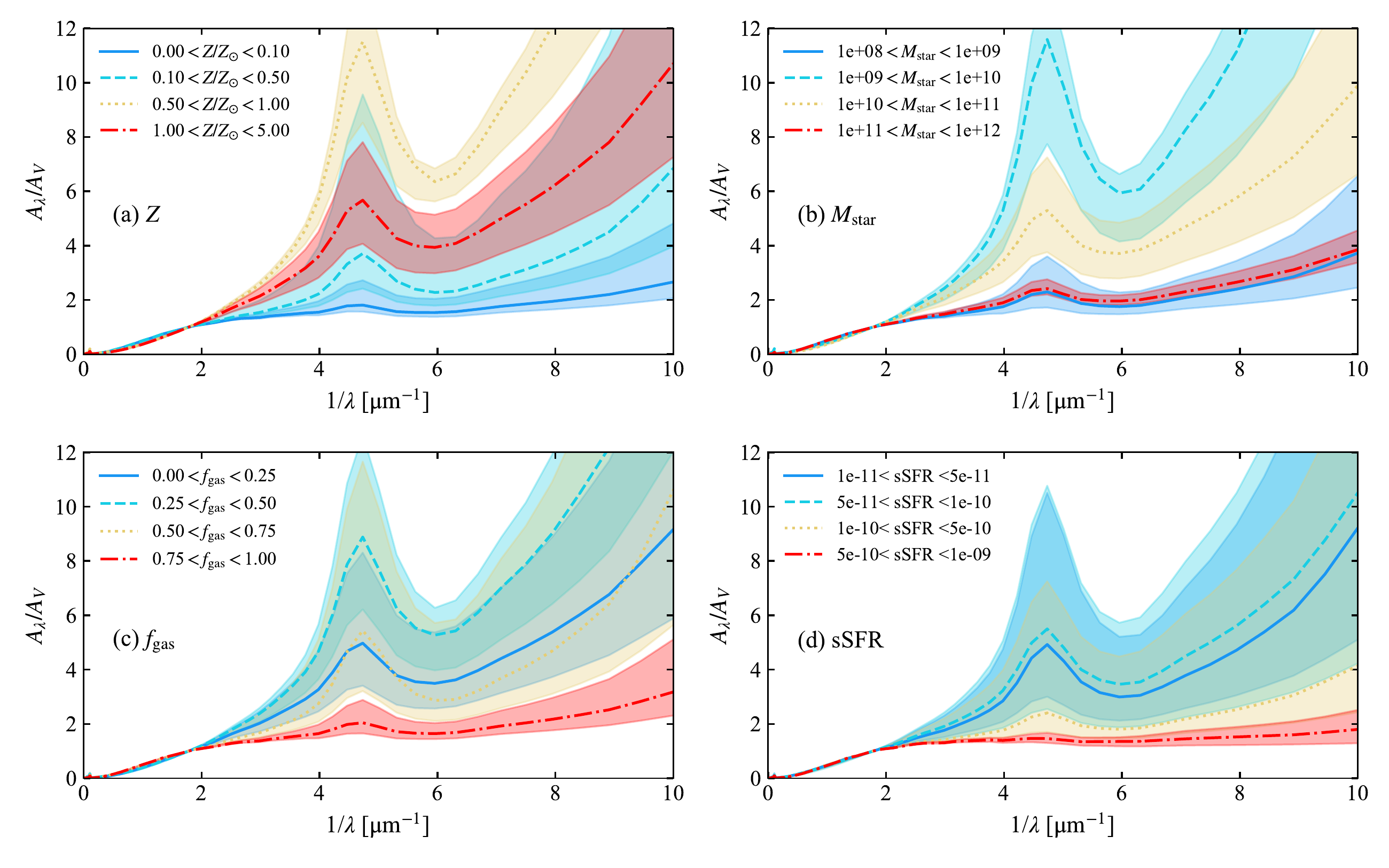}
  \caption{\label{fig:ext_dependence}
  Extinction curves binned by the metallicity (top left), stellar mass (top right), gas fraction (bottom left), and sSFR (bottom right).
  Each colour presents each bin as shown in the legend. The lines show the medians, while the shaded regions present the 25th and 75th percentile ranges.}
\end{figure*}

Fig.\ \ref{fig:ext_dependence} shows the extinction curves in four bins for metallicity, stellar mass, gas fraction and sSFR at $z=0$.
In general, the extinction curves are first steepened and then flattened as galaxies evolve.
This result is naturally understood from the grain size distributions shown in Fig.\ \ref{fig:size_dependence}.
The increase of small grains by accretion at the intermediate stage of galaxy evolution makes the extinction curves steeper, and coagulation, which subsequently convert small grains to large ones, makes them flatter.
Our model predicts that the extinction curve becomes the steepest in the intermediate stage of galaxy evolution.

We also note in Fig.~\ref{fig:ext_dependence} that
the 2175~\AA\ bump strength is always prominent except for the most `unevolved' bin of each quantity because of carbon enrichment by AGB stars is already efficient. 
This means that a steep extinction curve is associated with a strong 2175~\AA\ bump in our model. 
This is the main reason why the SMC extinction curve (a steep and bumpless curve) is difficult to reproduce with our model (Section \ref{subsec:ext_nearby}). 
Some physical mechanisms that could suppress small carbonaceous grains (e.g.\ selective loss of small carbonaceous grains; \citealt{Bekki:2015}) may need to be included to resolve this problem.

\begin{figure*}
  \includegraphics[width=2\columnwidth]{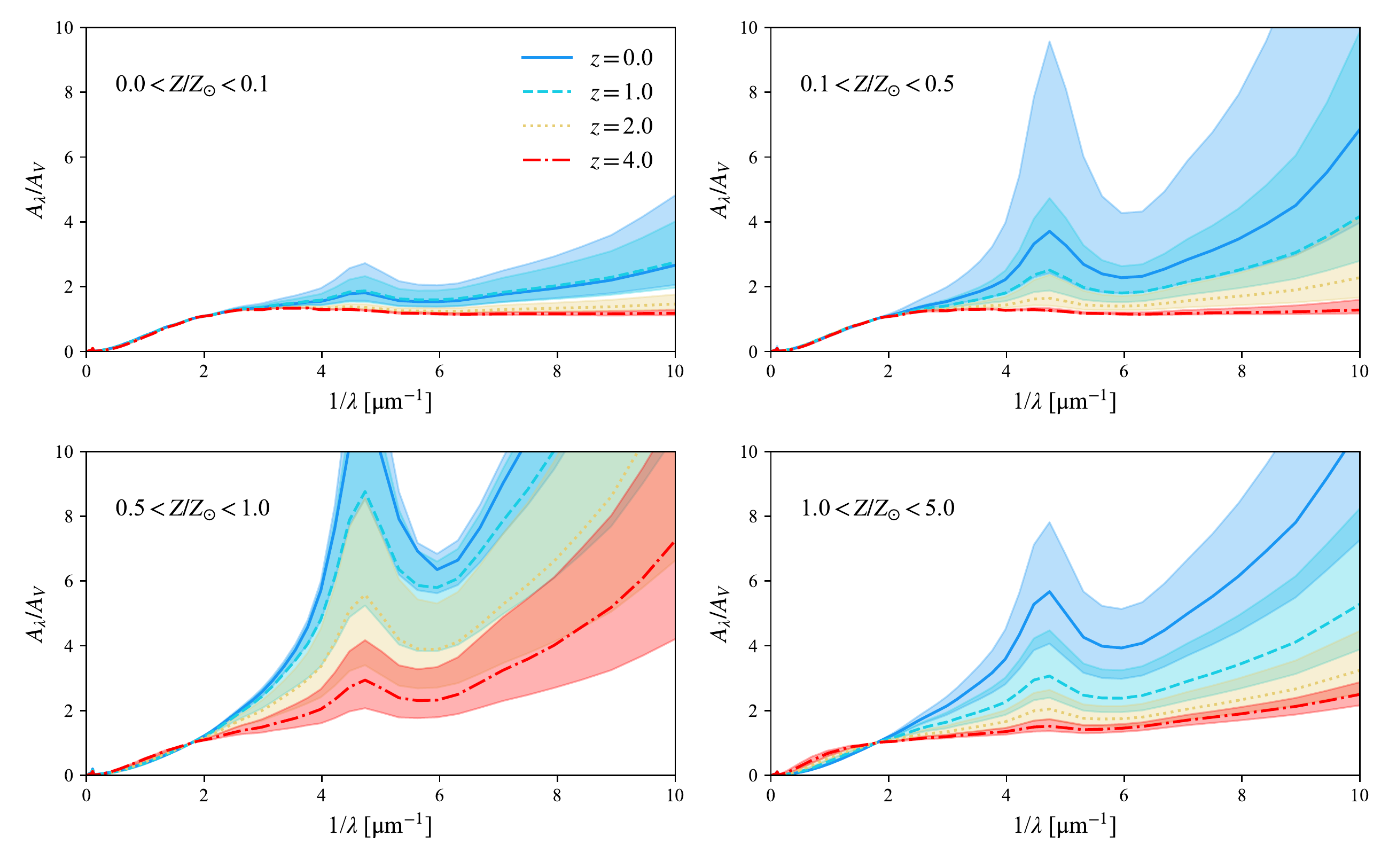}
  \caption{\label{fig:ext_z}
  Redshift evolution of dust extinction curves binned by the metallicity (shown in the upper left corner in each panel).
  The lines show the median and the shaded regions present the 25th and 75th percentile ranges. Each colour correspond to each redshift as shown in the legend.}
\end{figure*}

We also investigate the redshift evolution of dust extinction curves in four metallicity bins in Fig.\ \ref{fig:ext_z}.
In each metallicity bin, low-redshift galaxies have steeper extinction curves as expected from the increase of small grains in the grain size distribution shown in Fig.\ \ref{fig:size_z} (Section \ref{subsec:gsd}). 
From Fig.\ \ref{fig:ext_z}, we also observe that the metallicity dependence of extinction curves at each redshift is described by the change from flat to steep shapes at low to intermediate metallicities, and from steep to intermediate steepness at intermediate to high metallicities.

At high metallicity ($Z>1$ Z$_{\sun}$), the extinction curves at high redshift ($z\sim 4$) are very flat. 
This is due to an abundant micron-sized grains (see Fig.\ \ref{fig:size_z}), which are created by coagulation. 
Efficient coagulation reflects a high dense-gas fraction, which is linked to high star formation activity in our model (equation \ref{eq:dense}). 
Efficient star formation and lower dense gas fraction lead to an enhanced production of Si and a lower graphite fraction, and less efficient shattering. 
These effects make the 2175 \AA\ bump strength weak. 
Therefore, we predict that supersolar-metallicity objects at high redshift tend to have flat extinction curves with weak 2175 \AA\ bump strength.

\section{Summary and Conclusions}\label{sec:summary}

We implement the evolution of interstellar dust in a semi-analytic galaxy formation model ($\nu^2$GC), focusing on the evolution of grain size distribution.
We adopt the evolution model of grain size distribution developed by \citetalias{Hirashita/Aoyama:2019} and \citetalias{HM20} \citep[originally from][]{Asano/etal:2013}, who included stellar dust production, SN dust destruction, shattering, accretion and coagulation.
These processes are treated in a manner consistent with the star formation history, the metal enrichment, and the dense gas fraction.
Based on the calculated grain size distribution, we also predict the extinction curve as an observable characteristic of dust. 
The calculated grain size distribution for each galaxy is decomposed into silicate and carbonaceous species based on the sSFR and metallicity, and the latter species is further divided into graphite and amorphous carbon based on the dense gas fraction.
As a result of the above modelling, we are able to predict the statistical properties of dust masses, grain size distributions, and extinction curves for star-forming galaxies across cosmic history, together with the relation with other galaxy properties (metallicity, stellar mass, gas mass fraction, and sSFR).
Thus, this paper provides the most extensive prediction for the grain size distributions in both local and distant galaxies.

We successfully reproduce the dust properties in the nearby Universe in the following aspects: The relation between dust-to-gas ratio ($\mathcal{D}$) and metallicity ($Z$) at $z=0$ is consistent with that observed for nearby galaxies. 
We also confirm that the grain size distributions of the MW-like sample have shapes similar to the \citetalias{MRN} distribution. 
As a consequence, our model reproduces the MW extinction curve for the MW-like sample.
We also examine the total dust masses in galaxies and compared them with observed dust mass functions at various redshifts ($z\sim 0$--2.5). 
Our model overproduces the dust mass function for dusty galaxies at $z\lesssim 0.8$, while it is consistent with the observations at higher redshifts. 
The overprediction at low redshift is likely due to an overestimate of metallicity in massive galaxies. 
We tend to overproduce the cosmic dust mass density at all redshifts, but the discrepancy is within a factor $\sim 2$.
Both our model and observations show a peak of the comoving dust mass density at $z\sim 1$--1.5; thus, the qualitative behaviour of the cosmic dust mass evolution is reproduced well.

We finally show the relation between the grain size distribution and galaxy properties (metallicity, stellar mass, gas fraction, and sSFR). 
These quantities broadly reflect the evolutionary stages of galaxies. 
We observe that the grain size distribution shows the following evolutionary trend: The grain size distribution is dominated by large grains in the early stage (because of stellar dust production), it has a prominent overabundance of small grains (because of shattering and accretion) after that, and it finally converges to an MRN-like shape (becuase of coagulation).
This result also means that if we know one of the four quantities, we are able to infer the grain size distribution.
Note that the sSFR is less strongly correlated with the grain size distribution than the other three quantities.
As a consequence of the above evolutionary trend in the grain size distribution, the extinction curve shapes are sensitive to all those evolutionary parameters, such that they are flat, steep, and intermediate (MW-like) from the unevolved to evolved phases.
Our model predicts a strong correlation between the steepness and the 2175 \AA\ bump strength in extinction curves. 
This means that steep, bumpless extinction curves (like the SMC extinction curve) remain to be explained.

We also present the redshift evolution of grain size distributions binned by the metallicity. 
In a fixed metallicity range, the grain size distributions tend to have larger fractions of small grains with decreasing redshift. 
As a consequence, the extinction curves are steeper at lower redshift at a fixed metallicity.
The metallicity dependence of extinction curve at a fixed redshift has a similar trend to that at $z=0$.
We also find that supersolar-metallicity objects at high redshift tend to have flat extinction curves with weak 2175 \AA\ bump strength.

The `upgraded' $\nu^2$GC semi-analytic model developed in this paper will be useful to calculate not only the dust abundances of galaxies but also the grain size distributions and their related quantities (especially extinction curves) at various cosmic epochs. 
It would also be interesting to extend the predictions to attenuation curves \citep[e.g.][]{Calzetti:2001} with a proper treatment of radiative transfer \citep[e.g.][]{Narayanan:2018,Lin:2021} or to dust emission SEDs in the IR with an appropriate modelling of interstellar radiation field intensity \citep[e.g.][]{Relano:2020,Chang:2022,Nishida:2022}.
This step is useful for comparison with observational data at various redshifts \citep[e.g.][]{Aoyama:2019,Lim:2020}.
Since both attenuation curves and dust emission SEDs are largely affected by the grain size distribution, our framework is useful for predicting dust-related radiative properties of galaxies in the future.

\section*{Acknowledgements}
We thank the anonymous referee for useful comments.
We are grateful to Taira Oogi for fruitful discussions and useful comments and to the $\nu^2$GC collaboration members for continuous support.
RM is supported by JSPS KAKENHI Grant Number 20K14515.
HH thanks the National Science and Technology Council (NSTC) for support through grants 108-2112-M-001-007-MY3 and 111-2112-M-001-038-MY3, and the Academia Sinica for Investigator Award AS-IA-109-M02.

\section*{Data Availability}

Data related to this publication and its figures are available on request from the corresponding author.


\bibliographystyle{mnras}
\bibliography{draft} 


\appendix

\section{Silicate fraction}\label{app:fsil}

We explain how we derived the fitting formula for the silicate fraction ($f_\mathrm{sil}$) presented in equation (\ref{eq:fsil}). 
\citetalias{HM20} modelled the evolution of $f_\mathrm{sil}$ based on the abundance ratio between Si and C. 
They assumed exponentially decaying star formation histories with a wide varieties of decaying time-scales ($\tau_\mathrm{SF}=0.5$, 5, and 50 Gyr). 
We use these calculation results (see \citetalias{HM20} for details) to derive the silicate fraction in our model, in which 
an older age is characterized by a smaller sSFR. 
We expect that the above wide range of $\tau_\mathrm{SF}$ reflects some representative cases for the relation between the age and sSFR. 
We should, however, keep in mind that this correspondence is more complicated if we consider more complex star formation histories.

The silicate fraction is large at young ages (high sSFR) because of the dominant contribution of SNe to the metal enrichment, and is lowered later because of carbon production by AGB stars. 
However, it is difficult to describe $f_\mathrm{sil}$ only with a function of sSFR since an old age with long $\tau_\mathrm{SF}$ and a young age with short $\tau_\mathrm{SF}$ are not distinguished. 
We find that the situation is improved if we divide sSFR by the metallicity $Z$. 
With the same value of sSFR, a low (high) metallicity indicates that the stellar mass has (has not) been built up significantly in the past star formation so that SNe (AGB stars) tend to be the dominant source of the metal enrichment.

\begin{figure}
  \includegraphics[bb = 0 56 595 481, width=1\columnwidth]{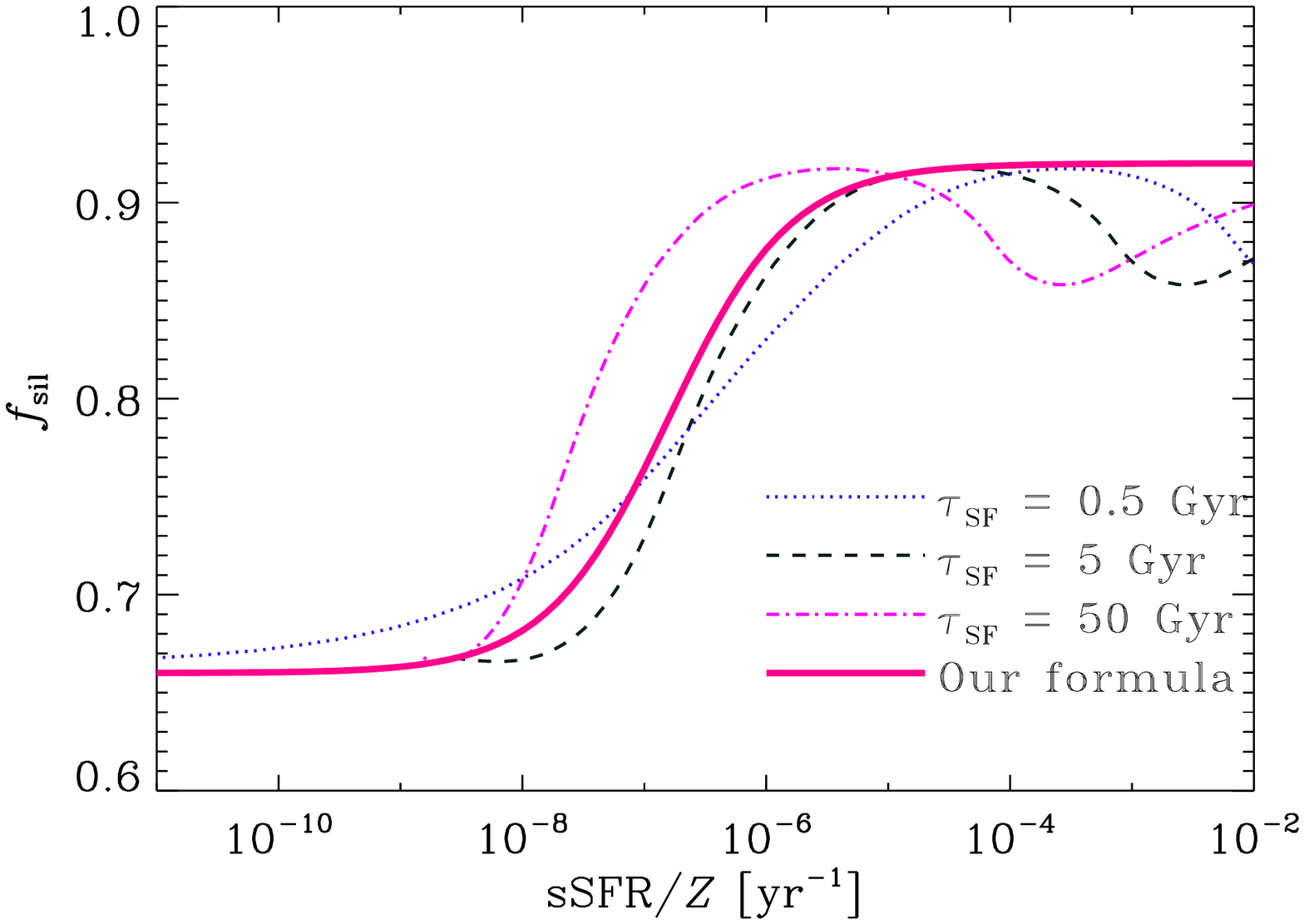}
  \caption{Silicate fraction as a function of $\mathrm{sSFR}/Z$. We use the value of sSFR in units of yr$^{-1}$ and the absolute metallicity (not normalized to solar) for $Z$. The dotted, dashed, and dot--dashed lines present the results with $\tau_\mathrm{SF}=0.5$, 5, and 50 Gyr, respectively. The solid line is the relation adopted in the text (equation \ref{eq:fsil}). Note that the galaxy evolves from large to small $\mathrm{sSFR}/Z$ in our model.
  \label{fig:fsil}}
\end{figure}

In Fig.\ \ref{fig:fsil}, we plot $f_\mathrm{sil}$ as a function of $\mathrm{sSFR}/Z$. 
We express sSFR in units of yr$^{-1}$ and $Z$ in absolute metallicity (i.e.\ not normalized to solar). 
We observe that the above three star formation histories with a wide variety in $\tau_\mathrm{SF}$ predict similar behaviour of $f_\mathrm{sil}$ as a function of $\mathrm{sSFR}/Z$. 
The transition of $f_\mathrm{sil}$ from large to small values happens around $\mathrm{sSFR}/Z\sim 10^{-7}$ yr$^{-1}$. 
Thus, we propose a fitting formula described in the text (equation \ref{eq:fsil}), which broadly reproduces $f_\mathrm{sil}$ of the three cases within an error of $\lesssim 0.05$.

\bsp    
\label{lastpage}
\end{document}